\documentclass[preprint]{revtex4-1}
\usepackage{graphicx}
\usepackage{amssymb}
\usepackage{multirow}
\usepackage{xcolor}

\begin{document}
\title {Curvature-dependent adsorption of water inside and outside armchair carbon nanotubes}
\author{Shulai Lei}
\affiliation{Institut f\"ur Chemie und Biochemie, Freie Universit\"{a}t Berlin, Takustra{\ss}e 3, 
D-14195 Berlin, Germany}
\author{Shujuan Li}
\affiliation{ Institut f\"ur Mathematik, Freie Universit\"{a}t Berlin, Arnimallee 6, D-14195 
Berlin, Germany}
\author{Burkhard Schmidt}
\affiliation{ Institut f\"ur Mathematik, Freie Universit\"{a}t Berlin, Arnimallee 6, D-14195 
Berlin, Germany}
\author{Beate Paulus}
\affiliation{Institut f\"ur Chemie und Biochemie, Freie Universit\"{a}t Berlin, Takustra{\ss}e 3,
D-14195 Berlin, Germany}
\begin{abstract}
The curvature dependence of the physisorption properties of a water molecule inside and outside an 
armchair carbon nanotube (CNTs) is investigated by an incremental density--fitting local coupled 
cluster treatment with single and double excitations and perturbative triples (DF--LCCSD(T)) study. 
Our results show that a water molecule outside and inside $(n,n)$ CNTs ($n$=4, 5, 6, 7, 8, 10) is 
stabilized by electron correlation. The adsorption energy of water inside CNTs decreases quickly 
with the decrease of curvature (increase of radius) and the configuration with the oxygen pointing 
towards the CNT wall is the most stable one. However, when the water molecule is adsorbed outside 
the CNT, the adsorption energy varies only slightly with the curvature and the configuration with 
hydrogens pointing towards the CNT wall is the most stable one. We also use the DF--LCCSD(T) results 
to parametrize Lennard--Jones (LJ) force fields for the interaction of water both with the inner and 
outer sides of CNTs and with graphene representing the zero curvature limit. It is not possible to 
reproduce all DF--LCCSD(T) results for water inside and outside CNTs of different curvature by a 
single set of LJ parameters, but two sets have to be used instead. Each of the two resulting sets 
can reproduce three out of four minima of the effective potential curves reasonably well. These LJ 
models are then used to calculate the water adsorption energies of larger CNTs, approaching the 
graphene limit, thus bridging the gap between CNTs of increasing radius and flat graphene sheets.
\end{abstract}

\maketitle

\section{Introduction}
In recent years, the study of water confined inside low--diameter carbon nanotubes (CNTs) has 
attracted a lot of attention because of many exotic properties that differ from those of the bulk 
phases \citep{Marti2001, Marti2001b, Marti2001a, Wu2011, jp4025206, ja5041539}. Being confined in 
tubes with diameters not much larger than the size of the water molecules themselves, the topology 
of their hydrogen--bonding networks is dominated by unique structures such as water wires, different 
helix--like or prism--like ice nanotubes, both single-- or multi--walled \citep{Gordillo2000, 
tasis2006chemistry, alexiadis2008molecular}. Consequently, there is also a variety of different 
phase transitions of water inside small CNTs not seen in bulk water, and in some cases it was 
predicted that ice nanotubes can even exist at room temperature \citep{kyakuno2011confined, 
Takaiwa2008}. Moreover, spontaneous electric polarization was found to occur in some ice nanotubes 
confined inside CNTs \cite{Luo2008, Li2015, nakamura2011ferroelectric}. Also the transport of water 
is strongly effected by the confinement in nanotubes. In particular, the greatly enhanced 
flow--rates are assigned to the smoothness of the CNT interior walls \citep{falk2010molecular}. It 
is expected that these properties may give rise to applications of CNTs for, e.g., biomedical 
devices and efficient molecular separation \citep{Ohba2014, Kou2014}.

As long as microscopic experiments concerning the structure and dynamics of water confined in small 
CNTs are still very scarce, most of the work in the fields mentioned above is based on molecular 
simulations. Because of the large number of particles typically involved in molecular dynamics (MD) 
simulations of water in CNTs, practically all previous work was based on empirical force fields. In 
particular, Lennard--Jones (LJ) potentials are widely used to model the non--bonded carbon--water 
interactions and are available in most of the molecular dynamics software packages, providing an 
efficient way to simulate systems even with millions of atoms. However, there are large 
discrepancies in the values of the empirical LJ parameters used in previous studies of structure and 
dynamics of water confined in CNTs, see e.g. Tab.~\uppercase\expandafter{\romannumeral1}  in Ref. 
\cite{alexiadis2008molecular}. But already in one of the first simulation studies of water 
conduction in CNTs, it has been noticed that small changes in modelling the nanotube--water 
interaction can induce large differences in the water occupancy \cite{Hummer2001} and water filling 
kinetics and thermodynamics \citep{Waghe2002, Vaitheeswaran2004, Sriraman2005, Waghe2012}. Also 
structure and dynamics of ice nanotubes confined inside CNTs were found to depend sensitively on the 
parameters of LJ potentials \citep{Perez-Hernandez2013, Li2015}. Hence, it is a major challenge to 
obtain quantitatively correct values for those parameters. With the recent advances in computational 
resources, paralleled by progress in the developments of methods and algorithms, high--level quantum 
chemical methods have become available, even for rather complex systems, such as the ones of 
interest here. Recently, the interaction of water with graphene has become a benchmark system for 
high--level quantum chemistry methods \cite{Cabaleiro-Lago2009, Jenness2009, Jenness2010, Rubes2009, 
Rubes2010}. In fact, incremental CCSD(T) results for water adsorbed on graphene 
\citep{Voloshina2011b} are used to reparametrize LJ interaction models for MD simulations 
\citep{Perez-Hernandez2013} of water confined inside CNTs. There an overall water--carbon 
interaction strength has been found, which is significantly stronger than in most previous 
simulations \cite{Koga2000a, Hummer2001, Koga2001, Berezhkovskii2002, Noon2002, Mashl2003, 
Kolesnikov2004, Wang2004, Bai2006, Alexiadis2008d, Kofinger2008, Luo2008, Takaiwa2008}. Moreover, 
the results for the water--carbon interaction showed a rather strong anisotropy with respect to the 
water orientation, which was not considered in most of the previous work either, for an exception 
see Ref. \cite{Kaukonen2012}.

The present work aims at accounting for the effect of the curvature of CNT walls on the interaction 
with water which has an important impact, e.g., on the transport properties \cite{falk2010molecular, 
Liu2005a}. However, to the best of our knowledge, until recently there have been no high level 
\textit{ab initio} results for the interaction of water with CNT systems but only density functional 
theory (DFT) results \cite{Kaukonen2012,rajarajeswari2011effect}. Hence, we lately started to 
investigate these systems at a highly accurate quantum--chemistry level, namely an incremental 
density-fitting local coupled cluster treatment with single and double excitations and perturbative 
triples (DF--LCCSD(T)) \cite{BeateShulai}, investigating the curvature dependent adsorption 
properties of a water molecule inside and outside $(n,n)$ CNT ($n$=4, 5, 6, 7, 8, 10) fragments. In 
order to save computational effort and to be able to treat even larger systems, the focus of this 
work is on a comparison of the computationally very demanding DF--LCCSD(T) method with different, 
less expensive methods: a) the computationally cheaper dispersion corrected DFT variants which are, 
however, not systematically improvable as the CCSD(T) method, and b) even cheaper LJ force fields 
for classical molecular dynamics simulations for larger CNTs, thus bridging the gap between CNTs of 
increasing radius and flat graphene sheets.

\section{Model and computational details }
In order to explore the adsorption interaction between CNTs and water, the system is modelled by a 
single water molecule inside or outside $(n,n)$ CNT fragments ($n$= 4, 5, 6, 7, 8, 10), where the 
ends are saturated by 4$n$ hydrogen atoms. The length of the CNT 
fragment (defined as distance between the two outer carbon rings perpendicular to the tube axis) in 
our simulation is  7.37 \AA{} in all cases, which results in 14$n$ carbon atoms for our $(n,n)$ CNT 
fragments. Based on the most stable configuration of water on a graphene surface 
\citep{Voloshina2011b}, throughout this work we keep the water molecule to be located above the 
center site of a carbon ring and the two hydrogen atoms of the water molecule pointing towards C--C 
bonds along the axial direction of CNTs. In each case, two configurations are investigated, as 
displayed in Fig.~1. One of them is termed  $\wedge$--configuration, in which the two hydrogen atoms 
are closer to the wall of CNT than the oxygen atom. The other configuration is termed 
$\vee$--configuration, in which the oxygen atom is closer to the wall of CNT. Due to 
the high computational costs of the quantum chemistry methods discussed below, scans of water 
orientation angles are beyond our computational means. For the water molecule, the O--H bond 
lengths are set to be 0.958 \AA{} and the H--O--H angle is set to be 104.45$^\circ$. For armchair 
CNTs, the C--C distances are set to the experimental value of 1.421 \AA{}.  The dangling bonds of 
CNTs are saturated with hydrogen atoms with C--H distances set to 1.084 \AA{}.

Highly accurate quantum--chemical CCSD(T) results using MOLPRO software \citep{MOLPRO_brief} are 
obtained using the method of increments \citep{stoll1992correlation}. In order to reduce the 
computational effort with only a negligible effect on the accuracy for large systems, the 
DF--LCCSD(T) method are used \citep{werner2011efficient}. The basis sets of the polarized 
correlation-consistent valence-double-$\zeta$ basis (cc--pVDZ) for C and H of CNTs and the 
aug--cc--pVTZ basis \citep{Dunning1989} for the O and H atom of water molecule are used for 
Hartree--Fock (HF) calculations and the corresponding basis for density--fitting for the local 
correlation treatments. A basis set test performed for water on graphene \citep{Voloshina2011b} 
shows that the used basis set is sufficiently accurate for the adsorption energy.

For comparison we also employ dispersion corrected density functional methods which are less 
time--consuming than CCSD(T) calculation. The PBE--D2/D3 calculations \citep{Perdew1996, Grimme2006, 
Grimme2010} are performed by using MOLPRO software \citep{MOLPRO_brief}  for the finite fragments 
and with projector augmented wave pseudopotentials as implemented in the Vienna ab--initio 
simulation package (VASP) \citep{Kresse1996a, Kresse1996} for a periodic set--up. In the projector 
augmented wave method, the plane--wave kinetic energy cutoff is set to 400 eV. The Brillouin--zone 
integration is sampled with a  single $k$--point. The CNT systems are modelled by using a 
(1$\times$1$\times$5) supercell to insure that the distance between a water molecule and its 
periodic image is more than 12 \AA{} to avoid notable interaction between them.  A comparison of 
the finite fragment PBE--D3 calculations (with the finite basis set (MOLPRO)) with the periodic 
modelling (with the plane-wave basis (VASP)) of the water inside and outside the CNT  yields very 
good agreement, providing that the cluster model of the CNTs is sufficiently larger. The shape of 
the potential energy curves and the positions of the minima agree very well, only the depth of all 
minima is deeper by about 20 meV in the plane-wave basis treatment.

Both for larger CNTs or for an increasing number of water molecules, even the computational effort 
of DFT calculations will be exceedingly high, at least for long--time molecular dynamics simulations. 
As an alternative, we aim here at obtaining reliable force field parameters by fitting effective 
water--CNT and water--graphene potentials to DF--LCCSD(T) results. To this end, we use the 
approximation of pairwise additive LJ potential energy functions which are often used to model 
non--bonded interactions:
\begin{equation}
V(r_{ij})=4\sum_{i \in {\rm H}_2{\rm O} \atop j \in {\rm CNT}}\epsilon_{ij} 
\left[\left(\frac{\sigma_{ij}}{r_{ij}}\right)^{12}-\left(\frac{\sigma_{ij}}{r_{ij}}\right)^6\right], 
\label{eq:LJ}
\end{equation}
which sums the interactions between H and O atoms ($i$) of the water model and 
C atoms ($j$) in the CNT (or graphene) model. Here $\epsilon_{ij}$ represents 
the respective attractive well depth and $\sigma_{ij}$ is the root of the pair potential, i.e. the 
point where attractive and repulsive part of the LJ potential cancel. Note that our 
LJ potentials are also used to implicitly model the CNT polarization by the water dipole moment; 
hence no partial charges are involved throughout the present work. As will be shown in detail below, 
the $\sigma$ parameters are very close to the sums at the respective van der Waals radii and none or 
only minor adjustments are necessary. In contrast, optimal values for the $\epsilon$ parameters are 
under debate in the literature \citep{Perez-Hernandez2013}. Hence, we scanned the range of those 
parameters manually, striving for good agreement with the current CCSD(T) results. 
An automated fitting procedure yielding also statistics of the deviations such as, 
e.g. in Ref. \citep{ruangpornvisuti1987interaction}, is not employed here because of the very small 
number of parameters (see Sec. \uppercase\expandafter{\romannumeral3} for details) and because of 
the relatively small data sets obtained from the quantum chemical calculations.

\section{Results and Discussion}
To quantify the curvature effect, we calculate the adsorption energies of a water molecule inside 
and outside of armchair $(n,n)$ CNT fragments ($n$=4, 5, 6, 7, 8, 10). The position of the center 
of mass of the water molecule which respect to the longitudinal axis of the $(n,n)$ CNT is labelled 
with \textit{x}. Negative values correspond to the $\wedge$--configuration outside CNTs and 
$\vee$--configuration inside CNTs, while positive values for $\wedge$--configuration inside CNTs and 
$\vee$--configuration outside CNTs, as shown in Fig.~1(a) from left to right.

Our DF-LCCSD(T) results are compiled in Tab. \uppercase\expandafter{\romannumeral1}. 
It is immediately seen that the adsorption energies of water inside CNTs are much stronger than 
those for water outside CNTs, because in the former case the attractive dispersive forces from the 
all sides of the nanotube contribute to the interaction energy add up \citep{BeateShulai}. In the 
later case, however, the water molecule essentially only interacts with one side of the CNT.
When the water molecule adsorbed outside $(n,n)$ CNTs (see Fig~2(a) and (d) for $n$=5,  Fig.~2(e) 
and (h) for  $n$=6, Tab.~\uppercase\expandafter{\romannumeral1}  for all cases), the DF--LCCSD(T) 
adsorption energy of the $\wedge$--configuration is always lower than the one of the 
$\vee$--configuration. For the smallest armchair CNT (4,4) investigated, a DF--LCCSD(T) adsorption 
energy of $\vee$--configuration of water outside is -99 meV, while -136 meV for the 
$\wedge$--configuration (not shown in Fig. 2). The adsorption energy curves for a water molecule 
outside $(6,6)$ CNT are very similar to those for water outside $(4,4)$ and $(5,5)$ CNTs and show 
nearly no curvature effects. Note that the adsorption energies of water outside $(n,n)$ CNTs , see 
Tab.~\uppercase\expandafter{\romannumeral1} also for larger CNTs, are close to the results of a 
water molecule adsorbed on a graphene surface, including the energy differences between $\wedge$-- 
and $\vee$--configurations which are -123 and -108 meV, respectively \citep{Voloshina2011b}.

Next, we will discuss three typical cases of a water molecule inside CNTs. First, water inside 
$(4,4)$ CNT shows purely repulsive interactions indicating that water molecules do not permeate 
spontaneously, since a $(4,4)$ CNT has a too small tube radius, $R$=2.65 \AA{}, to accommodate water 
molecules inside. Second, for $n$=5 the energy curves have only one minimum for water inside CNT 
corresponding to a $\vee$--configuration, as shown in Fig.~2(b). Therefore, this is the narrowest 
armchair CNT to adsorb water molecules inside. The third typical case is $n$=6. As shown in Fig.~2(f) 
and 2(g), two  minimum energy structures are predicted inside a (6,6) CNT by our DF--LCCSD(T) 
calculations. Also for larger diameter CNTs we find two minima at the DF--LCCSD(T) level of theory, 
in all cases with the $\vee$--configuration slightly more stable than the $\wedge$--configuration. 
An important result is that the investigated CNTs display an obvious curvature effect on water 
adsorbed inside them as listed in Tab.~\uppercase\expandafter{\romannumeral1}. The adsorption energy 
decreases quickly with decreasing wall curvature (increasing CNT radius $R$) from $n$=5 to 10 
approaching the graphene limit, $1/R\rightarrow 0$ (see open symbols in Fig.~\ref{fig:Limit}). While 
the adsorption energy of water inside a (5,5) CNT is nearly four times as large as the outside case, 
for the (10,10) CNT the energies of water inside and outside become close to each other. A change in 
the order of the stabilities of $\vee$-- and $\wedge$--configuration has to occur for large diameter 
CNTs, because for the adsorption on graphene the $\wedge$--configuration is more stable.

As also shown in Tab.~\uppercase\expandafter{\romannumeral1}, the stable configurations predicted by 
using MP2 are the same as the DF--LCCSD(T) for a water molecule adsorbed inside CNTs. The MP2 
calculations also reproduce the same tendencies of the adsorption curves but with a little higher 
energies than the DF--LCCSD(T) results. However, the VASP results for PBE--D2 and PBE--D3 are 
different for inside and outside cases. For the outside cases, PBE--D3 method gives a better 
description of dispersion correlation than PBE--D2. For the inside case, where the DF--LCCSD(T) 
calculations report two pronounced minima for $n\ge6$ and the $\vee$--configuration as the more 
stable one, both PBE--D2 and PBE--D3 predict only the $\wedge$--configuration as the stable one.

Another goal of the present work is to use the DF--LCCSD(T) results to parametrize force fields for 
the interaction of a single water molecule with graphene as well with CNTs 
(inside and outside). First, we consider our previous LJ model for the CO and CH
interaction between oxygen and hydrogen atoms in the water molecule and all of the carbon atoms, 
obtained by fitting to CCSD(T) calculations for the water--graphene interaction only 
\cite{Voloshina2011b, Perez-Hernandez2013}, shown in the first row (\#~1) of Tab.~\ref{tab:FF}. As 
can be seen from the blue curves in Fig.~\ref{fig:HO}, this model cannot reproduce our DF--LCCSD(T) 
results for the CNTs. In general, the interaction strength is weaker than predicted by DF--LCCSD(T), 
especially for water outside the CNTs. 
Upon considerably increasing the attractive well depth 
$\epsilon_\mathrm{CH}$ 
between carbon and hydrogen and slightly readjusting the two $\sigma$ parameters, (row \#~2 in 
Tab.~\ref{tab:FF} 
and red curves in Fig.~\ref{fig:HO}), the potential energy curves for water adsorbed outside CNTs are 
reproduced satisfactorily. 
However, the curves for water inside CNTs are considerably more attractive than the DF--CCSD(T) 
results, especially when hydrogen is pointing towards the CNT wall ($\wedge$--configuration). Hence, 
the energetic ordering of $\vee$-- and $\wedge$--configurations is exchanged, as was also found for 
parameter set \#~1. Our Fig.~\ref{fig:HO} shows also results for water--graphene systems, where (A), 
(B) and (C) are labelling configurations where the water molecule is on top of an atom, on a bond 
site or a ring center site, respectively. For case (A), set \#~2 describes the adsorption energy 
better than set \#~1; conversely, for case (B) and (C), parameter set \#~1 is superior.

To solve the problem of the energetic ordering of $\vee$-- and $\wedge$--configurations for water 
inside CNTs, we suggest a modified LJ model for the water--carbon interaction where the oxygen 
interaction site is replaced by two 
neutral dummy particles (labelled as X in Tab. \uppercase\expandafter{\romannumeral2}) representing the 
polarizability of the lone pair electrons of water, using the same positions as in the TIP5P force 
field \cite{Mahoney2000}. The model with the lone pair dummies stabilizes the $\vee$--configuration 
inside the CNTs due to a better description of the polarizability of the oxygen atom,  since the main 
contribution to the binding is the induced dipole--dipole interaction of the water molecule with the 
$\pi$--electron density of the CNTs \cite{BeateShulai}, and therefore the description of the 
polarizability of the oxygen atom was improved in the LJ model.  It can be seen in Fig.~\ref{fig:HD} 
that the new model describes the potential energy curves much better, especially for the 
$\vee$--configuration of water inside ($n,n$) CNTs. Two sets of parameters obtained 
by manually scanning the $\epsilon$ and $\sigma$ parameters for the CH and CX interaction are 
listed as \#~3 and \#~4 in Tab.~\ref{tab:FF} and shown as blue and red curves, respectively, in 
Fig.~\ref{fig:HD}. These two sets of LJ parameters have different advantages and disadvantages in 
describing the potential energy in the different cases. Parameter set \#~3 describes the potential 
energy inside CNTs everywhere very well and the $\vee$--configuration (but not the 
$\wedge$--configuration) for water outside CNTs. In contrast, parameter set \#~4 describes the 
potential energy outside CNTs everywhere very well and also the $\vee$--configuration (but not the 
$\wedge$--configuration) for water inside CNTs. Note that, within the restrictions of the simple LJ 
model, it was not possible to find one set of parameters describing the four minima for each of the 
CNTs equally well. Hence, in practical MD simulations, one of these two sets of parameters can be 
chosen, depending on the different confinement conditions of water molecules. In describing the 
water--graphene adsorption energy, parameter set \#~4 is better than parameter \#~3 for sites (A) 
and vice versa for sites (B) and (C).

The resulting LJ models can be used to predict also the interaction of water with CNTs of much 
larger radius which are currently out of reach for high level quantum chemistry. Calculated data for 
parameter sets \#~3 and \#~4 for armchair CNTs ($n,n$) with $n$=12, 14, 16, 18, 22, 30, 34, 38, 42 
are also included as full symbols in Fig.~\ref{fig:Limit}. Parameter set \#~3 reasonably reproduces 
the magnitude of the water--graphene interaction for the inside adsorption minima, but the order of 
the $\wedge$-- and $\vee$--configuration is wrong in the limit, $1/R\rightarrow 0$. Using parameter 
set \#~4 for the extrapolation to the graphene limit, the $\wedge$--configuration is stabilized 
against the $\vee$--configuration, but too strong, so the magnitude of the adsorption energy is 
overestimated. However, set \#~4 can reproduce the transition from the most stable 
$\vee$--configuration inside small CNTs to the most stable $\wedge$--configuration for larger CNTs 
and graphene leading to a crossing of the corresponding curves. Parameter set \#~4 predicts a radius 
of about 10 \AA{} which corresponds to a (14,14) CNT for the crossing. For larger radii the 
$\wedge$--configuration is also the most stable one inside the CNTs.

\section{Summary and conclusions}
The physisorption properties of a water molecule interacting with $(n,n)$ CNTs with $n$=4, 5, 6, 7, 
8, 10 are investigated by performing DF--LCCSD(T) calculations. Our main result is that the 
physisorption properties of water molecule outside and inside carbon nanotube display a different 
curvature dependence of the adsorption energies. For the outside case, there is only little effect 
of the curvature, however, for the inside case, the magnitude of the adsorption energy is decreasing 
with decreasing CNT curvature of CNT wall, tending toward the graphene results. By fitting LJ 
parameters to our high quality DF--LCCSD(T) quantum chemistry results, we were able to derive simple 
but reliable models for carbon sheets of different curvature. In particular, by adding dummy 
particles mimicking the polarizability of the oxygen atom we improved the quality of atom--centered 
LJ models. This model reproduces all DF--LCCSD(T) results for water inside and outside CNTs of 
different curvature by two sets of LJ parameters. Each of them can reproduces three out of four 
minima of the effective potential curves reasonably well. These LJ models are used to calculate the 
water adsorption energies of larger CNTs, approaching the graphene limit, thus bridging the gap 
between CNTs of increasing radius and flat graphene sheets.
Finally, we mention that, in agreement with our previous findings for water interacting with graphene 
\citep{Perez-Hernandez2013, Voloshina2011b}, the overall carbon--water interaction strength obtained 
from fitting to the current DF--LCCSD(T) results for water interaction with CNTs is considerably 
stronger than assumed in most of the previous classical MD simulations \citep{Alexiadis2008d, 
alexiadis2008molecular}, with the exception of the work presented in Refs. \citep{Kaukonen2012} 
and \citep{jz2012319}. Hence, the present work suggests that the CNTs may be less hydrophobic than 
commonly anticipated.

\section{Acknowledgements}
We appreciate financial support from the Focus Area Nanoscale of the Free University of Berlin. 
The computer facility of the Free University of Berlin (ZEDAT) is acknowledged for computer time.  
Shujuan Li is grateful to the Chinese Scholarship Council for financial support. Helpful discussions 
with Dr. Krista Grace Steenbergen, Udbhav Ojha, Matthias Berg and Dr. Carsten M\"uller are acknowledged.
\bibliographystyle{rsc}
\bibliography{bib}
\newpage
\begin{table*}[t]\renewcommand{\arraystretch}{1.1} 
\caption{\label{tab:table1} Adsorption energies (in meV) of a water molecule outside  and inside 
$(n,n)$ CNT with $n$=5, 6, 7, 8, 10 for $\wedge$--configurations and $\vee$--configurations for 
various quantum chemistry methods. The values in parentheses are minimum energy positions (in \AA{}) 
along $x$ axis as shown in Fig. 1(a). The PBE--D2/D3 are from periodic calculations using VASP 
\citep{Perdew1996, Grimme2006, Grimme2010, Kresse1996, Kresse1996a}.}
\begin{ruledtabular}
\begin{tabular}{lll ccccc}
\multicolumn{2}{c}{Configurations}              &Methods     &(5,5)      &(6,6)      &(7,7)      &(8,8)      &(10,10)    \\ 
\hline
\multirow{8}*{Outside} &\multirow{4}*{$\wedge$} &DF--LCCSD(T) &-126(-6.6) &-128(-7.3) &-128(-7.9) &-127(-8.7) &-124(-10.1)  \\
                       &                        &MP2         &-117(-6.6) &-117(-7.3) &-118(-8.0) &-115(-8.7) &-110(-10.1)  \\
                       &                        &PBE--D2      &-130(-6.4) &-131(-7.1) &-133(-7.8) &-134(-8.5) &-136(-9.9)  \\
                       &                        &PBE--D3      &-124(-6.6) &-124(-7.3) &-127(-8.0) &-128(-8.7) &-131(-10.1)  \\
\cline{2-8}                            
                       &\multirow{4}*{ $\vee$}  &DF--LCCSD(T) &-103(6.1)  &-103(6.8)  &-104(7.5)  &-101(8.2)   &-96(9.6)   \\
                       &                        &MP2         &-86(6.1)   &-87(6.9)   &-88(7.6)   &-85(8.3)    &-79(9.6)   \\
                       &                        &PBE--D2      &-82(6.3)   &-81(7.0)   &-83(7.7)   &-82(8.4)    &-84(9.8)   \\
                       &                        &PBE--D3      &-88(6.4)   &-89(7.1)   &-91(7.8)   &-91(8.5)    &-93(9.9)   \\
\cline{1-8}
\multirow{8}*{ Inside} &\multirow{4}*{$\wedge$} &DF--LCCSD(T) &-404(0.0)  &-262(0.5)  &-199(1.4) &-171(2.1) &-145(3.5)  \\      
                       &                        &MP2         &           &-235(0.5)  &-176(1.3) &-150(2.1) &-134(3.5)  \\
                       &                        &PBE--D2      &-357(0.0)  &-282(0.8)  &-237(1.5) &-213(2.3) &-190(3.7)  \\
                       &                        &PBE--D3      &           &-312(0.5)  &-249(1.3) &-219(2.1) &-192(3.5)  \\
\cline{2-8}
                       &\multirow{4}*{$\vee$}   &DF--LCCSD(T) &           &-279(-0.9) &-214(-1.7) &-182(-2.5)  &-148(-3.9)  \\
                       &                        &MP2         &-364(-0.1) &-257(-0.9) &-197(-1.7) &-166(-2.4)  &-143(-3.9)  \\
                       &                        &PBE--D2      &           &           &-167(-1.5) &-144(-2.3)  &-121(-3.8)  \\
                       &                        &PBE--D3      &-378(-0.1) &           &-202(-1.5) &-169(-2.3)  &-142(-3.7)  \\        
\end{tabular}
\end{ruledtabular}
\end{table*}
\newpage

\begin{table}
\begin{tabular}{c c c c c c c}
\hline\hline
 $\#$      &$\epsilon_\mathrm{CO}$ & $\epsilon_\mathrm{CH}$    &$\epsilon_\mathrm{CX}$ & $\sigma_\mathrm{CO}$  &$\sigma_\mathrm{CH}$    &$\sigma_\mathrm{CX}$ \\
\hline
1  &5.182    &2.591    &             & 3.16             &2.73               & \\
\hline
2  &5.182    &4.457    &             & 3.00             &2.80               & \\
\hline
3  &         &3.835    &3.6274       &                  &2.85               &2.60 \\
\hline
4  &         &6.633    &2.178        &                  &2.75               &2.65 \\
\hline\hline
\end{tabular}
\caption{LJ potential parameters for water--carbon interaction, see Eq. (\ref{eq:LJ}), where 
$\epsilon$ represents the attractive well depth (in meV) 
and $\sigma$ is the distance at which the pair potential is zero (in \AA{}). Subscripts C, O, H and 
X are for carbon, oxygen, hydrogen and dummy particles 
of water respectively (no partial charges involved). Row \#~1: previous fit to CCSD(T) calculations 
of graphene--water from Ref. \cite{Perez-Hernandez2013}. \#~2--4: fit to the present DF--LCCSD(T) 
calculations of ($n,n$) CNT--water for $n=5, 6, 7, 8, 10$.}
\label{tab:FF}
\end{table}
\newpage

\clearpage
\begin{figure}[!htp] 
\includegraphics[width=10cm]{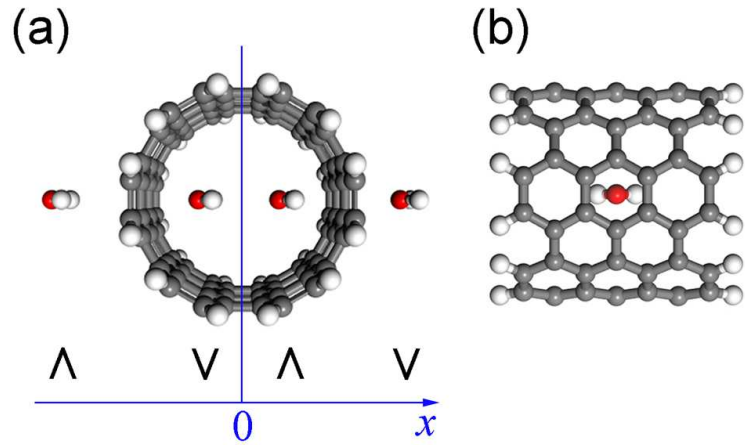}
\caption{\label{str} Adsorption structures of a water molecule inside and outside a (6,6) CNT 
fragment. (a) top view for $\vee$--configuration outside, $\wedge$--configuration inside, 
$\vee$--configuration inside and $\wedge$--configuration outside CNT (from left to right) and (b) 
side view.}
\end{figure}
\newpage

\begin{figure}[t] 
\includegraphics[width=\textwidth]{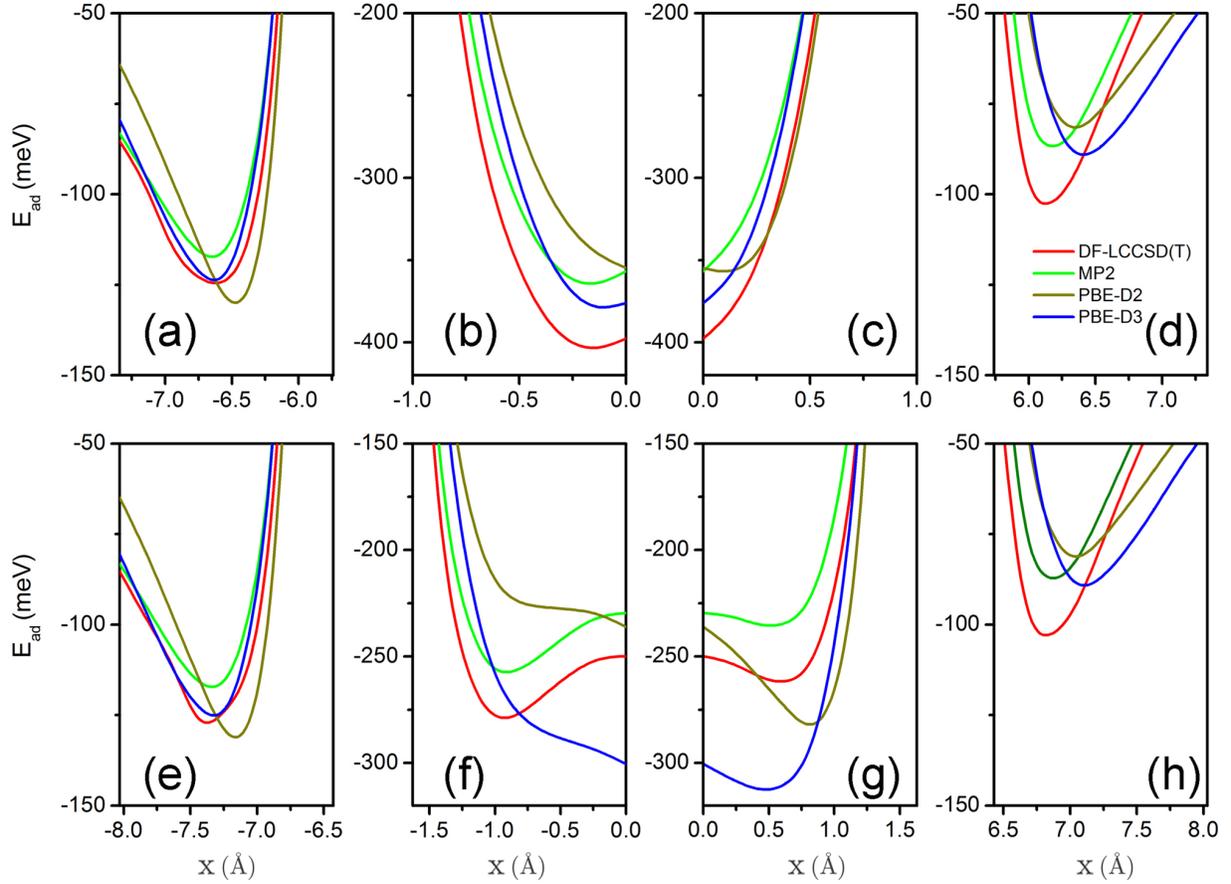}
\caption{\label{55} Adsorption energies of water inside and outside CNTs for different methods as a 
function of the position $x$. (a) $\wedge$ outside, (b) $\vee$ inside, (c) $\wedge$ inside and (d) 
$\vee$ outside (5,5) CNT. (e)$\sim$(h) same for (6,6) CNT. The PBE--D2/D3 are from periodic 
calculations using VASP \citep{Perdew1996, Grimme2006, Grimme2010, Kresse1996, Kresse1996a}.}
\end{figure}
\newpage

\begin{figure}
\centering           
\includegraphics[width=10cm]{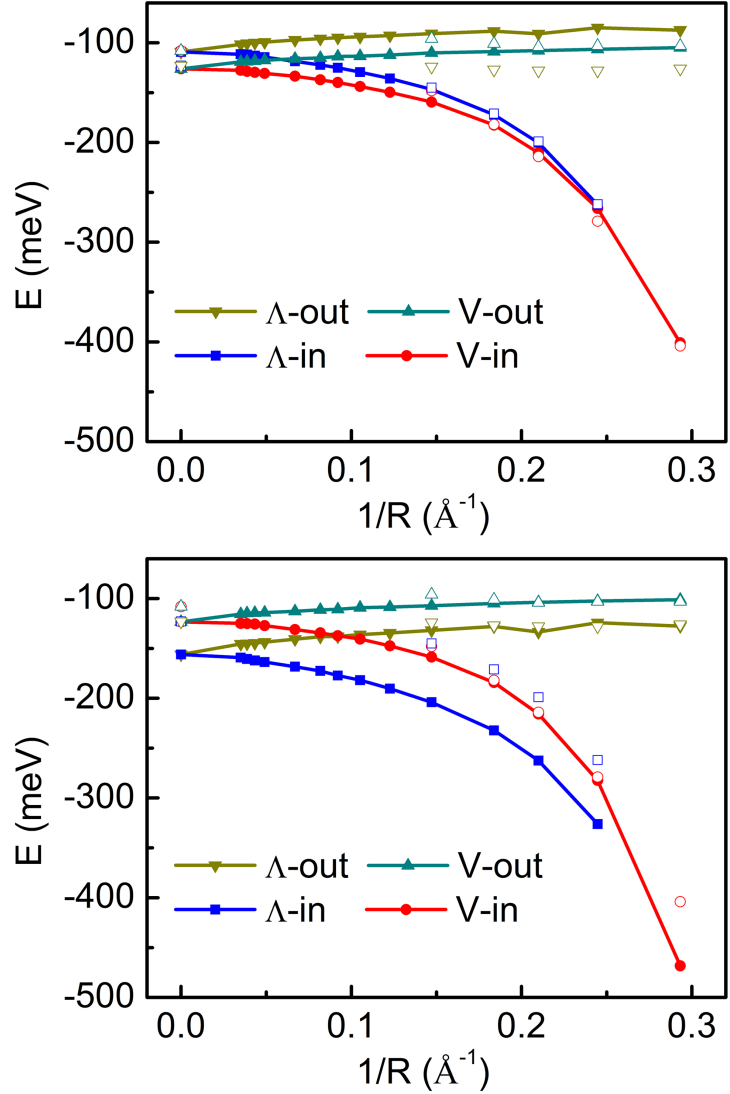}
\caption {Curvature dependence of the four minima of the effective potential energy curves of water 
inside and outside CNTs. Open symbols: DF--LCCSD(T) results for CNTs, (see also Tab. 
\uppercase\expandafter{\romannumeral1}) and CCSD(T) results for graphene, $1/R\rightarrow 0$, from 
Ref. \cite{Voloshina2011b}. Full symbols with curves: LJ model with parameter set \#~3 (upper part) 
and \#~4 (lower part), also Tab. \uppercase\expandafter{\romannumeral2}}
\label{fig:Limit}
\end{figure}
\newpage

\begin{figure}
\centering           
\includegraphics[width=\textwidth]{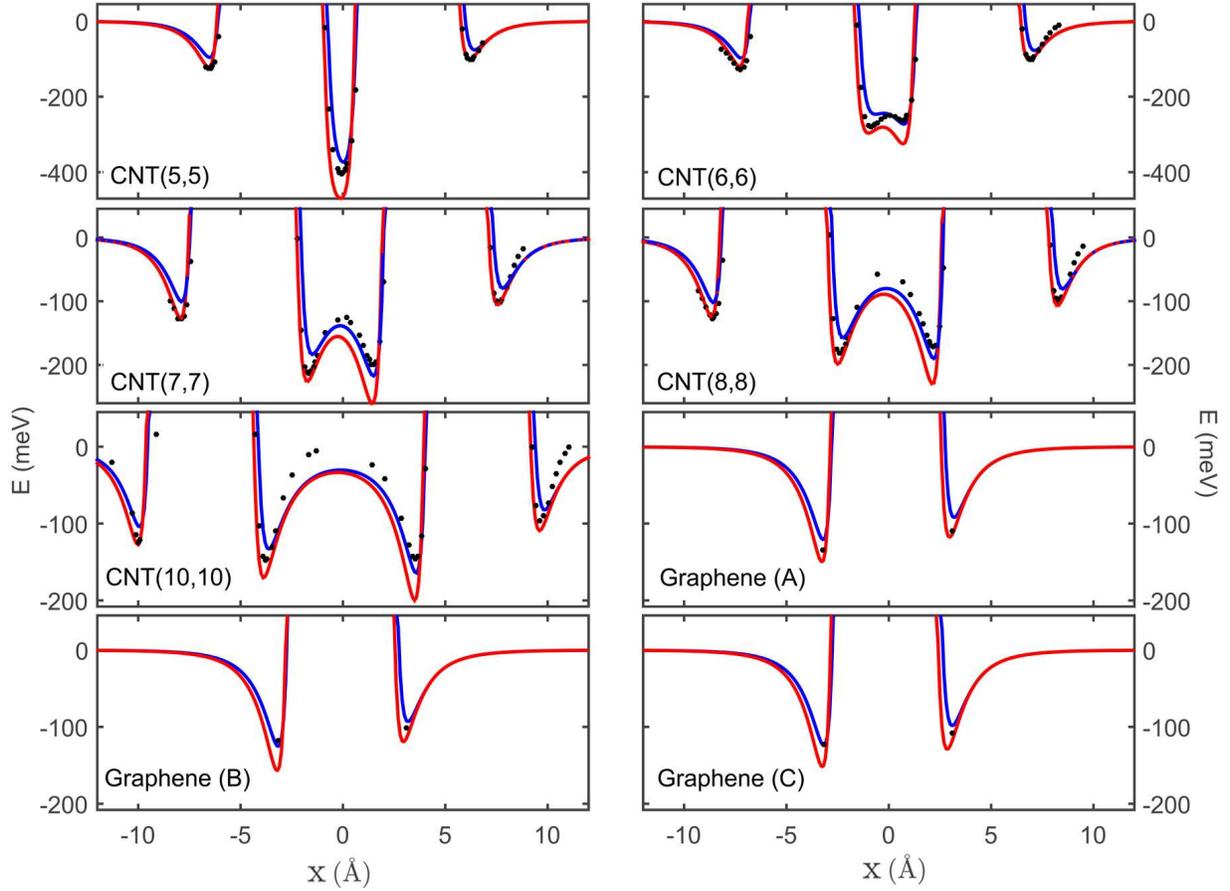}
\caption{Scan of the potential energy of a water molecule traversing various CNTs and graphene. 
Force fields \#~1 from previous work \cite{Perez-Hernandez2013} (blue) and \#~2 (red) with 
interaction sites for water O and H, for parameters see Tab.~\ref{tab:FF}. Black points represent 
DF--LCCSD(T) results of the present work for CNTs and CCSD(T) results of previous work 
\cite{Voloshina2011b} for graphene.}
\label{fig:HO}
\end{figure}
\newpage

\begin{figure}
\centering           
\includegraphics[width=\textwidth]{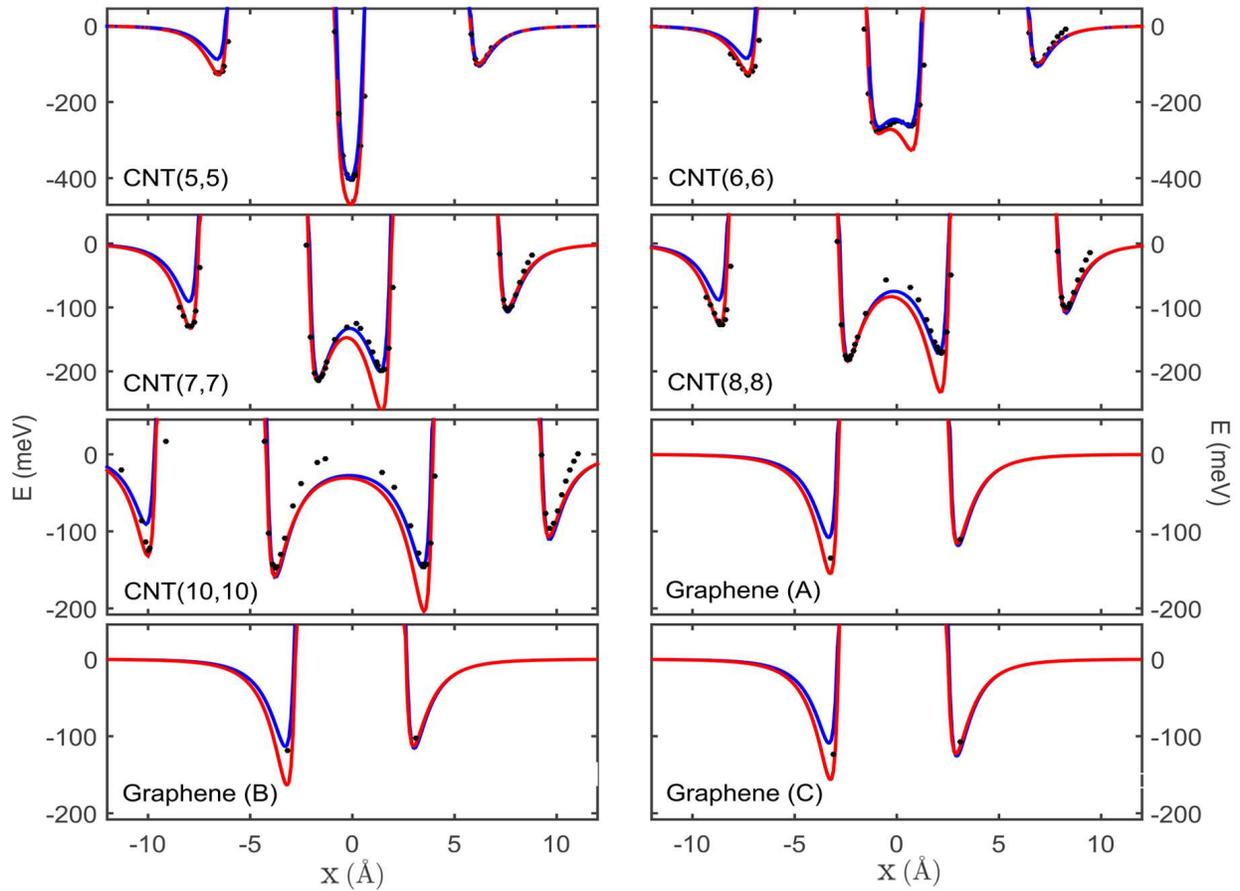}
\caption {Same as Fig. \ref{fig:HO} but for a modified LJ model with additional dummy particles 
representing the lone pairs of the water molecule. Force fields \#~3 (blue) and \#~4 (red), for 
parameters see Tab.~\ref{tab:FF}.}
\label{fig:HD}
\end{figure}
\end{document}